\documentclass[10pt]{article}
\usepackage{amssymb}
\usepackage{amsmath}
\usepackage{wasysym}
\usepackage{graphicx}
\usepackage{appendix}
\date{}
\begin{document}

\title{Induced osmotic vorticity in the quantum hydrodynamical picture}

\author{C. Dedes\thanks{{c\_dedes@yahoo.com}}\\
Bradford College  \\
Great Horton Road, Bradford \\
West Yorkshire, BD7 1AY \\
United Kingdom \\}

 \maketitle

\begin{abstract}
A nonlinear wave mechanical equation is proposed by inserting an imaginary quantum potential into the Schr\"{o}dinger equation. An explicit expression for its solution is given under certain assumptions and it is shown that it entails attenuation related effects as non-unitary evolution, non-exponential quantum decay and entropy production. In the quantum hydrodynamical formulation the existence of circulation effects for the osmotic velocity field is established. Finally, a time-invariant equation for the probability density is derived, analogous to the tensor Lighthill equation in aeroacoustics, which admits both retarded and advanced solutions.

\end{abstract}

\section{Introduction}

One of the most spectacular aspects of quantum theory is that it can be formulated in a variety of ways that even appear not to be equivalent at first sight. In some occasions the mathematical formulation is identical but the conceptual framework differs significantly, in other instances even the mathematical formalism varies. It is well-known that soon after the inception of quantum mechanics de Broglie proposed an alternate formulation based on a guiding wave field, an idea rediscovered later by Bohm \cite{Bohm52}. Almost at the same time it became possible to formulate the new theory in terms of a hydrodynamical model \cite{Madelung} and this line of enquiry was later continued and perfected by Takabayasi \cite{Takabayasi}. In both formulations, although in a different context, a prominent place is given to what is called quantum potential. The Bohm quantum potential should not be considered as another classical potential function acting on physical systems \cite{Sanz}. It is a rich physical concept with broader mereological implications and displays with clarity the inherent wholeness and non-separability of an interacting many-body system \cite{Bohm} since it does not necessarily diminish with distance and it is manifestly non-local. On the other hand its very reliance on a probabilistic measure, namely the curvature of the amplitude of the wavefunction \cite{Sanz}, seems problematic within the scope of a deterministic theory like the de Broglie-Bohm one. In the next section we examine some fundamental considerations, introducing a non-linear wave mechanical equation and relate its solution with the linear Schr\"{o}dinger equation. In section 3 we examine the hydrodynamic formulation and examine the effect of the quantum potential to the osmotic velocity. A time symmetrical wave equation with a non-homogeneous term for the probability density is also derived that applies to the unmodified dynamical problem.

\section{Foundational issues}

Here we modify the Schr\"{o}dinger dynamics by introducing an imaginary Bohmian potential and then make a polar substitution for the wavefunction. In that way we have introduced a quantum potential term in both the Hamilton Jacobi and the modified continuity equation as a point of contact and consequently we recombine those two expressions in a different manner assuming at the same time that the phase of the nonlinear wavefunction is approximately the same with that of the linear equation. Our point of departure is a modified nonlinear wave equation of the form

\begin{equation}
i\hbar\frac{\partial \Phi}{\partial t} = 
 \left(-\frac{\hbar ^{2}}{2m}\nabla ^{2}+V(\mathbf{r},t)\right)\Phi -i\varepsilon  Q\Phi, 
\end{equation}

\noindent
where

\begin{equation}
    Q=-\frac{\hbar^{2}}{2m}\frac{\nabla^{2}R}{R}=-\frac{\hbar^{2}}{4m\rho}\left[\nabla^{2}\rho-\frac{(\nabla \rho)^{2}}{2\rho}\right]
\end{equation}

\noindent
the Bohm quantum potential discussed earlier. Making a standard polar substitution of the form $\Phi=Re^{iS/\hbar}$ and $S$ the multivalued phase of the wavefunction gives the continuity equation

\begin{equation}
  \frac{\partial R}{\partial t}+\frac{1}{m}\nabla R\cdot\nabla S+\frac{1}{2m}R\nabla ^{2}S=\frac{\varepsilon\hbar}{2m}\nabla ^{2}R,
\end{equation}

\noindent
which can also be written as 

\begin{equation}
 \frac{\partial \rho}{\partial t}+\nabla
\cdot\left(\rho\frac{\nabla S}{m}\right)=-\frac{2\varepsilon}{\hbar}\rho Q . 
\end{equation}

\noindent
The Hamilton-Jacobi equation reads as
 
 \begin{equation}
     \frac{\partial S}{\partial t}=-\frac{1}{2m}\left({\nabla S}\right)^{2}-V(\mathbf{r},t)-Q.
 \end{equation}

\noindent
 It follows from (4) that even if the phase does not exhibit any spatial variation the probability density may still be time-dependent. It can be reduced to a nonlinear equation, transformed to a diffusion equation like (3) and be readily solved provided that $\varepsilon$ is positive, otherwise the coefficient of diffusion will be negative and the problem ill-posed. On the contrary $\varepsilon<0$ describes amplification effects. A diffusion equation requires initial conditions to be imposed in order to deal with a well-posed problem, in other words if we use final conditions for (3) or (4) this engenders the possibility of a blow-up to infinity and a singularity in a finite time interval. This means that there is a preferred time direction and it is not possible to retrodict solutions in the backward time direction. If we allow only the first order derivative phase term to remain we obtain a diffusion equation with convection that can still be solved exactly. The existence of the Bohmian potential in (1) allows us to establish a connection between the Hamilton-Jacobi and continuity equations as

\begin{equation}
 \frac{\partial \rho}{\partial t}+\nabla
\cdot\left(\rho{\frac{\nabla S}{m}}\right)=-\frac{2\varepsilon}{\hbar}  \rho\left[\frac{\partial S}{\partial t}+\frac{1}{2m}(\nabla S)^{2}+V(\mathbf{r},t)\right].
\end{equation}

\noindent
The right hand of the above expression, which can been seen as a source term for the continuity equation, is related to the classical Schr\"{o}dinger equation \cite{Holland}. The time-independent version of (6) is easily obtained through the relations $\frac{\partial \rho}{\partial t}=0$, $\frac{\partial S}{\partial t}=-E $. By introducing the convective derivative which is expressed as

\begin{equation}
    \frac{d}{dt}=\frac{\partial}{\partial t}+\mathbf{v}\cdot \nabla=\frac{\partial}{\partial t}+\frac{\partial}{\partial t}\frac{\nabla S}{m}\cdot \nabla,
\end{equation}

\noindent
it is possible to solve (6) for $\rho=R^{2}$

\begin{equation}
    R(\mathbf{r},t)=e^{\int_{t_{0}}^{t}\left[-\frac{2\varepsilon}{\hbar}\left(\frac{\partial S}{\partial \tau}+\frac{1}{2m}(\nabla S) ^{2}+V(\mathbf{r},\tau)\right)-\frac{\nabla^{2}S}{2m}\right]d\tau}R(\mathbf{r}_{0},t_{0}).
\end{equation}

\noindent
As in \cite{Wyatt} we can write

\begin{equation}
    e^{i\frac{S(\mathbf{r},t)}{\hbar}}=e^{\frac{i}{\hbar}\int_{t_{0}}^{t}L d \tau}e^{i\frac{S(\mathbf{r}_{0},t)}{\hbar}},
\end{equation}

\noindent
where $L$ the system Lagrangian. Multiplying the last two equations we obtain

\begin{subequations}
\begin{align}
  & \Phi(\mathbf{r},t)=e^{\int_{t_{0}}^{t}\left[-\frac{2\varepsilon}{\hbar}\left(\frac{\partial S}{\partial \tau}+\frac{1}{2m}(\nabla S) ^{2}+V(\mathbf{r},\tau)\right)-\frac{\nabla^{2}S}{2m}\right]d\tau}e^{\frac{i}{\hbar}\int_{t_{0}}^{t}L d \tau}\Phi(\mathbf{r}_{0},t_{0}), \\
  & \Phi(\mathbf{r},t)=e^{\int_{t_{0}}^{t}\left(-\frac{2\varepsilon}{\hbar}Q-\frac{\nabla^{2}S}{2m}\right)d\tau}e^{\frac{i}{\hbar}\int_{t_{0}}^{t}L d \tau}\Phi(\mathbf{r}_{0},t_{0}).
\end{align}
\end{subequations}

\noindent
In that way we have generalized the expression for the quantum hydrodynamic propagator given in the de Broglie-Bohm theory \cite{Wyatt}. Notice two alternative gauge-independent expressions are given. These are formulas relating the wavefunction $\Phi$ at two different times and spatial locations, through an exponential factor that involves integration over past times. It would be interesting to find a relation between $\Phi$ and $\Psi$ at the same time and location. In order to do this we need to make an assumption about the phase. We assume that the phase of the unperturbed system does not change significantly when we include the non-linear term. It is reasonable to assume that the imaginary term will more appreciably affect the evolution equation of the probability density and to a lesser extent the time development of the phase since the Hamilton-Jacobi that governs the evolution of the phase equation will remain unaltered. Certainly, we can add correction terms to this initial phase and use (5) to obtain iteratively better approximate expressions. Using (8) and setting $\varepsilon=0$ we find

\begin{equation}
A(\mathbf{r},t)=e^{-\frac{1}{2m}\int_{t_{0}}^{t}d\tau \nabla^{2}S}A(\mathbf{r},t_{0}),
\end{equation}

\noindent
where $\Psi=Ae^{iS/\hbar}$. Multiplying both sides with $e^{iS/\hbar}$ and assuming that  $R(\mathbf{r}_{0},t_{0})=A(\mathbf{r}_{0},t_{0})$ we find

\begin{subequations}
\begin{align}
  & \Phi(\mathbf{r},t)=e^{-\frac{2\varepsilon}{\hbar}\int_{t_{0}}^{t}\left[\frac{\partial S}{\partial \tau}+\frac{1}{2m}(\nabla S) ^{2}+V(\mathbf{r},\tau)\right]d\tau}\Psi (\mathbf{r},t), \\
  & \Phi(\mathbf{r},t)=e^{-\frac{2\varepsilon}{\hbar}\int_{t_{0}}^{t}Q_{0}d\tau}\Psi (\mathbf{r},t).
\end{align}
\end{subequations}

\noindent
Hence we have found a proportionality relation between $\Phi$ and $\Psi$. Notice that the quantum potential $Q_{0}$ corresponds to the state $\Psi$. This approximate expression indicates that the total probability amplitude is rescaled by an exponential integral factor with local phase dependency. When the finite duration of the time aperture becomes instantaneous so $t_{0}\rightarrow t$ the exponential becomes unity and we recover the linear solution $\Psi$ at equal times. Obviously we recover the linear limiting case too when $\varepsilon\rightarrow 0$. Within a many particle description the phase modifies a coincidence probability density and features the non-local character of the generated dissipation. It must be remembered that an essential hypothesis, apart from the inclusion of the imaginary Bohmian potential, was the assumption that the phase dependence of the modified wavefunction is identical to the linear Schr\"{o}dinger equation. Only then can we obtain this compact relationship between the two wavefunctions which would otherwise be unrelated. Since $\Phi$ is a solution of a nonlinear equation (1) in contradistinction to $\Psi$ it does not satisfy the superposition principle due to the existence of the scale factor. Nevertheless, it has the same phase so it may exhibit interference and correlation effects exactly as a coherent superposition of Schr\"{o}dinger equation solutions. A coherent state for example will display the same phase properties but an amplitude with a different spatio-temporal modulation. The presence of the state associated quantum phase illustrates the kind of indivisible, essential unity which is characteristic of the whole quantic system, more clearly pronounced in the multi-particle case. Normalization concerns dictate that $\Phi$ goes to zero at infinity and it may even involve non-Markovian evolution and entail memory effects \cite{Breuer}. Using (12) we can re-express (1) as a non-linear Schr\"{o}dinger equation with an added integral term. This is not a particular useful form but it is included here for the sake of completeness

\begin{equation}
i\hbar \left(\frac{\partial \Psi}{\partial t}-\frac{2\varepsilon}{\hbar} Q_{0}\Psi\right)=\left(\hat{T}+V\right)\Psi-\frac{2\varepsilon}{\hbar}\left(\int \hat{T}Q_{0}d\tau+Q \right)\Psi.
\end{equation}

\noindent
Effectively the existence of the imaginary Bohm potential means that the quantum action becomes complex

\begin{equation}
    S' \rightarrow S+{\frac{2i\varepsilon}{\hbar}\int_{t_{0}}^{t}\left[\frac{\partial S}{\partial \tau}+\frac{1}{2m}(\nabla S-q\mathbf{A}) ^{2}+V(\mathbf{r},\tau)\right]d\tau},
\end{equation}

\noindent
where a vector potential has been introduced in the usual manner as $\mathbf {v}=\frac{\nabla S}{m}-\frac{q}{m}\mathbf{A}$. The above relation suggests the following form for the non-relativistic propagator from place $\mathbf{r}_{0},t_{0}$ to $\mathbf{r},t$

\begin{equation}
  K=\int _{\mathbf{r}_{0}}^{\mathbf{r}} {\mathcal{D}}[\mathbf{r}(\tau)] e^{iS[\mathbf{r}(\tau)]-\frac{2\varepsilon}{\hbar} \int_{t_{0}}^{t}\left[\frac{\partial S[\mathbf{r}(\tau)]}{\partial \tau}+\frac{1}{2m}(\nabla S[\mathbf{r}(\tau)]-q\mathbf{A}) ^{2}+V(\mathbf{r},\tau)\right]d\tau}.
\end{equation}

\noindent
This expression clearly implies non-unitary evolution. Again it must be emphasized that it is this phase or action of the system that generates dissipation, so it is a kind of self-induced decay and unlike the influence functional or similar approaches we do not need to have recourse to a heat bath that affects the sub-dynamics of the system under examination. This is of great consequence and illustrates the sharp conceptual difference between the way irreversibility appears in the present formalism and more common reservoir-system approaches. In fact, it may be argued that strictly speaking it is not possible to trace out environmental degrees of freedom and divide rigorously a system from its surroundings, there is no sharp distinction between measured system and environment since the quantum phase cannot be appropriated to individual particles but points to the irreducible quantum compound and reveals its substantial unicity. If the action in the frequency domain is analytic in the upper half then its real and imaginary parts are not independent but they are related by a Hilbert integral transform

\begin{equation}
    \mathfrak{Re}[\Phi(\mathbf{r},\omega)] =\frac{1}{2\pi ^{2}}\mathcal{P}\int _{-\infty}^{+\infty}\frac{\mathfrak{Im}
    [\int _{-\infty}^{+\infty} g(\mathbf{r},\omega '-\omega)\Psi(\mathbf{r},\omega)d\omega]} {\omega-\omega '}d\omega,
\end{equation}

\noindent
where 

\begin{equation}
    g({\mathbf{r},\omega})=\frac{1}{2\pi}\int _{-\infty}^{+\infty}dt  e^{-\frac{2\varepsilon}{\hbar}\int_{t_{0}}^{t}\left(\frac{\partial S}{\partial \tau}+\frac{1}{2m}(\nabla S-q\mathbf{A})
    ^{2}+V(\mathbf{r},\tau)\right)d\tau} e^{i\omega t},
\end{equation}

\noindent
the frequency domain representation of the integral scale factor and $\Psi(\mathbf{r},\omega)$ that of $\Psi(\mathbf{r},t)$. Furthermore using the Fock-Krylov formula \cite{Ghirardi} for the survival probability of an unstable system we find

\begin{equation}
    p=\left|e^{-\frac{\varepsilon}{\hbar}\int_{t_{0}}^{t}\left[\frac{\partial S}{\partial \tau}+\frac{1}{2m}(\nabla S-q\mathbf{A}) ^{2}+V(\mathbf{r},\tau)\right]d\tau}\int _{-\infty}^{+\infty} g(E)e^{-iEt} dE\right|^{2},
\end{equation}

\noindent
which in general deviates from exponential decay law even if the density of energy states is a Lorentzian function . The decay rate is

\begin{equation}
\Lambda=\frac{2\varepsilon}{\hbar}\frac{\frac{\partial S}{\partial \tau}+\frac{1}{2m}(\nabla S-q\mathbf{A}) ^{2}+V(\mathbf{r},\tau)}{e^{-\frac{2\varepsilon}{\hbar}\int_{t_{0}}^{t}\left[\frac{\partial S}{\partial \tau}+\frac{1}{2m}(\nabla S-q\mathbf{A}) ^{2}+V(\mathbf{r},\tau)\right]d\tau}}.
\end{equation}

\noindent
 It is straightforward to generalize (1) for a many-particle system (of equal masses for convenience) or even a continuous one like a boson field. It is evident that if it is possible to decompose the phase as a sum of functions $S=S_{1}+S_{2}+..$ then the multiparticle wavefunction is separable. The rate of change of the total probability $ P=\int \rho d^{3}x$

\begin{equation}
    \frac{dP}{dt}=\int \frac{\partial \rho}{\partial t} d^{3}x=-\frac{\varepsilon\hbar}{2m}\int \frac{1}{2\rho}\sum _{i=1}^{N}\left(\nabla _{i} \rho\right)^{2}  d^{3}x < 0,
\end{equation}

\noindent
We see then that the total probability is not conserved but diminishes with time. What is important is that (1) entails the possibility of entropy production. Using the entropy formula $ S=-\int \rho ln\rho d^{3}x$ we obtain

\begin{equation}
    \frac{dS}{dt}=\frac{\varepsilon\hbar}{2m}\int \left(1+\frac{1}{2}ln\rho\right) \frac{1}{2\rho}\sum _{i=1}^{N}\left(\nabla _{i} \rho\right)^{2}  d^{3}x > 0,
\end{equation}

\noindent
which is positive. The positivity of entropy production is a general result following directly from the modified dynamics and does not rely on the phase approximation made earlier.

\section{Retrocausal solutions and osmotic vorticity}

It was stated in the introduction that quantum mechanics can be expressed in terms of quantum hydrodynamic variables, namely the probability density and the momentum field, even though the two formulations are not equivalent as certain hydrodynamical solutions may not satisfy the Schr\"{o}dinger equation. The inhomogeneous continuity equation for the probability density is analogous to the conservation of mass equation in fluid mechanics. It is possible to derive an equation analogous to the momentum Navier equation \cite{Takabayasi,Holland,Wyatt}. Employing (4) we find the continuity equation for the flux momentum 

\begin{equation}
    \frac{\partial (\rho \mathbf{v})}{\partial t}+\nabla \cdot ( \rho \mathbf{v}\mathbf{v})-\rho\frac{d\mathbf {v}}{dt}=-\frac{2\varepsilon}{m\hbar} \rho Q\mathbf{v},
\end{equation}

\noindent
where the quantum force is the sum of the classical and quantum potentials

\begin{equation}
\mathbf{F}=m\frac{d\mathbf v}{dt}=-\nabla (V+Q).
\end{equation}
 
\noindent
It is possible to write (22) in a tensor form as in \cite{Holland}

\begin{equation}
  \frac{\partial (\rho v_{i})}{\partial t}+\frac{\partial (\rho u_{i}v_{j})}{\partial x_{j}}=-\frac{\rho}{m} \frac{\partial V}{\partial x_{i}}-\frac{\partial\sigma_{ij}}{\partial x_{j}}-\frac{2\varepsilon}{m\hbar}\rho Q v_{i},
\end{equation}

\noindent
where the probability quantum stress tensor is written as

\begin{equation}
  \sigma _{ij}=\frac{\hbar ^{2}}{4m^{2}}\left(\partial_{i}\partial _{j}\rho-\partial _{i}\rho \partial _{j}ln\rho \right).
  \end{equation}

\noindent
We recover the familiar expression for the tensor equation \cite{Holland} in the limiting case $\varepsilon\rightarrow 0$. Following a development of the Navier-Stokes equations found in the aerodynamic sound literature \cite{Howe,Mattei,Ffowcs} we take the time derivative of the continuity equation for the probability density (4) and the divergence of (24) and by subtracting them we eliminate the convection term. In that way we combine the two in a new exact form as

\begin{equation}
   \frac{\partial ^{2}\rho}{\partial t ^{2}}+\frac{2\varepsilon}{m\hbar}\frac{\partial (\rho Q v_{i})}{\partial t} =\frac{\partial ^{2}}{\partial x_{i}\partial x_{j}}\left(\rho v_{i}v_{j}-\frac{\rho}{m} V\delta _{ij}+\sigma_{ij}\right),
   \end{equation}
   
\noindent
which contains both second and first order time derivatives and in order to find its time evolution we need to know initial and final conditions for $\rho$ and $\rho _{t}$ both. The imposition of two-time boundary conditions is already familiar but an important remark that needs to be made is that here the primary object of interest is not the quantum state but instead the probability function $\rho$ which we believe lays on a higher level of ontological density. The second term in (26) contains non-constant coefficients so the equation cannot be solved exactly but notice that it resembles the telegrapher's equation which admits solutions expressed by Heaviside step functions. This indicates at least qualitatively that the solutions of (26) will propagate not instantaneously but with finite speed. As in the aeroacoustics case we subtract from both sides a second spatial derivative term multiplied by an arbitrary reference velocity $c_{0}$ that states the propagation of $\rho$ and obtain 

\begin{equation}
   \frac{\partial ^{2}\rho}{\partial t ^{2}}+\frac{2\varepsilon}{m\hbar}\frac{\partial (\rho Q v_{i})}{\partial t} -c_{0} ^{2} \frac{\partial ^{2}\rho}{\partial x_{i}\partial x_{i} } =\frac{\partial ^{2}T_{ij}}{\partial x_{i}\partial x_{j}},
   \end{equation}
   
 \noindent
 where the driving Lighthill tensor which includes a quantum potential component with highly non-classical properties is expressed as
   
 \begin{equation}
       T_{ij}=\rho v_{i}v_{j}-\rho\left( \frac{V}{m}+c_{0}^{2}\right)\delta _{ij}+\sigma_{ij}.
 \end{equation}
   
 \noindent
 For a two-dimensional problem we could write (27) in terms of a stream function for the flow velocity $\mathbf{v}$. If we omit the quantum potential term and set $\varepsilon=0$ in (27) we obtain an evolution equation invariant under time reversal which admits solutions in both time directions. In contrast, the Schr\"{o}dinger equation admits solutions only in one direction in time and its complex conjugate in the opposite one. In that limit (27) is reduced to an inhomogeneous wave equation with a Lighthill source term \cite{Mattei,Ffowcs} and its solution is
   
  \begin{equation}
       \rho (\mathbf{x},t) = \frac{1}{4\pi c_{0} ^{2}}\frac{\partial ^{2}}{\partial x_{i}\partial x_{j}}\int _{-\infty}^{+\infty} \frac{T_{ij}\left(\mathbf{y},t-\frac{|\mathbf{x}-\mathbf{y}|}{c_{0}}\right)+ T_{ij}\left(\mathbf{y},t+\frac{|\mathbf{x}-\mathbf{y}|}{c_{0}}\right)}{|\mathbf{x}-\mathbf{y}|}d^{3}\mathbf{y},
   \end{equation}
   
   \noindent
 where $r=|\mathbf{x}-\mathbf{y}|$ the distance between a reception point $\mathbf{x}$ and the source $\mathbf{y}$. Even though this is a formal solution it still allow us to extract some significant conclusions by including both retarded and advanced terms. The existence of the retarded term is rather uncontroversial but this cannot be said for the advanced term since unavoidably introduces an element of finality and retrocausation \cite{Evans}. A similar issue is well-known in classical radiation theory where the advanced terms are excluded on physical grounds, but it is not clear if the same kind of argumentation could be employed here. Notice that in the quantum version of the single particle Lighthill equation derived here the quantum potential is embedded in the right hand side tensor. This implies nonlocal effects and strongly suggests that we cannot discard advanced solutions without loss of essential physics. Furthermore, for two or more particles the coincidence probability density existing in the 3N dimensional configuration space is connected to the total probability that appears in Bell type theorems which is a clear indication that the advanced term needs to be included in order to violate the classical correlations. This is obviously not a mathematical argument and the matter needs further consideration but it is plausible to assume that the effect of advanced terms in a many-particle context would definitely affect correlations and violate the usual assumed conditions for Bell-type theorems \cite{Price}. Obviously in the context of a classical electromagnetic or acoustic field there is no possibility of violating that family of inequalities but this is cerainly not the case for a quantum coincidence probability. One of course could escape the conclusion by modifying the ordinary Schr\"{o}dinger dynamics. Introducing now the stochastic osmotic velocity \cite{Wyatt,la Pena}, which is also implied in earlier formulas as in (21) and (25),

\begin{equation}
  \mathbf{u}=\frac{\hbar}{2m}\left(\frac{\nabla \Phi}{\Phi}+\frac{\nabla \Phi^{*}}{\Phi^{*}}\right)
\end{equation}

\noindent
gives

\begin{equation}
    \mathbf{u}=\frac{\hbar}{2m}\nabla ln |\Psi |^{2}-\frac{2\varepsilon}{m} \int_{t_{0}}^{t} \left(\frac{\partial  \mathbf{v}}{\partial \tau}+2\mathbf{v}\nabla \cdot\mathbf{v} +\nabla V\right)d\tau.
\end{equation}

\noindent
 This is a measurable, phase dependent diffusive velocity guided by the deterministic flow velocity field. The many particle expression exhibits dependency on the remote site locations which is a manifestation of nonlocality. Notice that integration over past times can be extended in principle to the initial probability distribution which makes a common cause explanation \cite{Dickson} implausible. The total velocity is the stochastic one in addition to the standard flux velocity. The intrinsic statistical natural of the osmotic velocity field does not allow deterministic trajectories and the rule of non-crossing is not applicable. The dilatation of the osmotic field is given by

\begin{equation}
\nabla\cdot \mathbf{u}=\frac{\hbar\nabla^{2}ln |\Psi |^{2}}{2m}-\frac{2\varepsilon}{m} \int_{t_{0}}^{t} \left(\frac{\partial  (\nabla\cdot\mathbf{v})}{\partial \tau}+2(\nabla\cdot\mathbf{v})^{2}+2\mathbf{v}\cdot\nabla ^{2} \mathbf{v} +\nabla^{2} V\right)d\tau.
\end{equation}

\noindent
 More interesting is the vorticity of this stochastic velocity field. Taking the curl of (31) we confirm that the first term is zero as the gradient of a scalar function but the integral term has non-zero contribution and some interesting implications. Using the fact that $\nabla\times\mathbf{v}=0$ yields 

\begin{equation}
   \boldsymbol{\omega}= \nabla \times \mathbf{u}=-\frac{4\varepsilon}{m} \int_{t_{0}}^{t} d\tau (\nabla^{2}\mathbf{v})\times \mathbf{v}.
\end{equation}

\noindent
We see then that the quantum velocity has a non-zero component related to circulation which could affect in principle the electron current in a conducting ring as in the case of mesoscopic Josephson currents. It also follows that other quantities related to circulation as the helicity $H=\int _{V} \mathbf{u} \cdot \boldsymbol{\omega} d^{3}x$ may be non-zero too. The osmotic acceleration is given by

\begin{equation}
  \frac{d\mathbf u}{dt} =\frac{d}{dt}\nabla  ln|\Psi |^{2}-\frac{2\varepsilon}{m}\left(\frac{\partial  \mathbf{v}}{\partial \tau}+2\mathbf{v}\nabla \cdot\mathbf{v} +\nabla V\right)
\end{equation}

\noindent
and the vorticity rate of change is

\begin{equation}
  \frac{d\boldsymbol {\omega}}{dt} =-\frac{4\varepsilon}{m}(\nabla^{2}\mathbf{v})\times \mathbf{v}.
\end{equation}

\noindent
Defining the circulation as

\begin{equation}
 \Gamma(t)=\oint _{C}  \mathbf{u}\cdot d \mathbf{r}=
 \iint _{\Sigma}(\nabla \times \mathbf{u}) \cdot d\mathbf{\Sigma}=-\frac{4\varepsilon}{m}\iint _{\Sigma} \left(\int_{t_{0}}^{t} d\tau (\nabla^{2}\mathbf{v})\times \mathbf{v}\right) \cdot d\mathbf{\Sigma},
\end{equation}

\noindent
and applying Stokes theorem we find

\begin{equation}
    \frac{d\Gamma(t)}{dt}=\oint _{C} \frac{d\mathbf{u}}{dt}\cdot d \mathbf{r}= -\frac{4\varepsilon}{m}\iint _{\Sigma}\left[\int_{t_{0}}^{t} d\tau (\nabla^{2}\mathbf{v})\times \mathbf{v}\right]\cdot d\mathbf{\Sigma}\neq  0.
\end{equation}

\noindent
Hence, the circulation of the osmotic field does change with time and the Kelvin circulation theorem \cite{Howe,Saffman} is violated. It is important to note that as with the entropy production and the tensor equation (27), the non-zero vorticity effect is again a general result following from the inclusion of an imaginary quantum potential and does not rely on the phase assumption mentioned in section 2. It can be immediately verified that the same result holds by substituting the exact expression (10) in (30).  

\section{Concluding remarks}

We have seen that the addition of an imaginary Bohm potential generates quantum friction and leads to certain interesting implications. It induces decay effects in the amplitude of probability, modifies the exponential decay law, leads to entropy production and generates vorticity effects for the osmotic velocity. The phase of the wavefunction, and its derivative velocity field, has particular importance in the present formalism since it guides the osmotic velocity field. Furthermore, still in the hydrodynamic formulation a generalized quantum hydrodynamic propagator has been obtained. In the domain of standard dynamics, by combining the two hydrodynamical equations a single non-homogeneous quantum Lighthill wave equation, manifestly time-symmetric, has been derived which in general admits both retarded and advanced solutions. The wider physical implications of this result suggest that we may need to choose between backward causation and exploring the possibility of extending the standard formalism beyond the unitary regime. 

\section*{Appendix}
\subsection*{Biparticle compound}
We consider a system of two spinless interacting particles with equal masses in one dimension ($\varepsilon=0$). Following the derivation in \cite{Wyatt} for the single particle case, we derive the probability density local continuity equation ($\rho=\rho _{12} (x_{1},x_{2})$,$S=S _{12} (x_{1},x_{2})$) 

\begin{equation}
    \frac{\partial \rho}{\partial t}+\frac{\partial }{\partial x_{1}}(\rho v_{1})+\frac{\partial }{\partial x_{2}}(\rho v_{2})=0
\end{equation}

\noindent
and the corresponding Hamilton-Jacobi one

 \begin{equation}
     \frac{\partial S}{\partial t}=-\frac{1}{2m}\left({\nabla _{1}S}\right)^{2}-\frac{1}{2m}\left({\nabla _{2}S}\right)^{2}-V-Q.
 \end{equation}
 
 \noindent
 Taking the appropriate spatial derivatives we find
 
 \begin{equation}
   \frac{\partial  v_{1}}{\partial t}= -v_{1}\frac{\partial v_{1}}{\partial x_{1}}-v_{2}\frac{\partial v_{2}}{\partial x_{1}}-\frac{1}{m_{1}}\frac{\partial}{\partial x_{1}}(V+Q)
 \end{equation}
 
  \begin{equation}
   \frac{\partial v_{2}}{\partial t}= -v_{1}\frac{\partial v_{1}}{\partial x_{2}}-v_{2}\frac{\partial v_{2}}{\partial x_{2}}-\frac{1}{m_{2}}\frac{\partial}{\partial x_{2}}(V+Q) 
 \end{equation}

 \noindent
 Using the formula
 
 \begin{equation}
   \frac{\partial (\rho v_{1,2})}{\partial t}=  \frac{\partial \rho}{\partial t}v_{1,2}+\rho \frac{\partial v_{1,2}}{\partial t}
 \end{equation}

\noindent
we obtain these Navier-Stokes equations for the momentum fields

 \begin{equation}
   \frac{\partial (\rho v_{1})}{\partial t}=-\frac{\partial (\rho v^{2}_{1})}{\partial x_{1}}-\frac{\rho}{m_{1}} \frac{\partial (V+Q)}{\partial x_{1}}-u_{1}\frac{\partial (\rho v_{2})}{\partial x_{2}}-\rho v_{2}\frac{\partial v_{2}}{\partial x_{1}}
 \end{equation}
 
  \begin{equation}
   \frac{\partial (\rho v_{2})}{\partial t}=-\frac{\partial (\rho v^{2}_{2})}{\partial x_{2}}-\frac{\rho}{m_{2}} \frac{\partial (V+Q)}{\partial x_{2}}-u_{2}\frac{\partial (\rho v_{1})}{\partial x_{1}}-\rho v_{1}\frac{\partial v_{1}}{\partial x_{2}} 
 \end{equation}
 
 \noindent
 As earlier we differentiate (38) over time and, (43) over $x_{1}$ and (44) over $x_{2}$ and subtract from the first the sum of the two latter. Using the relation
 
 \begin{equation}
     \frac{\partial v_{1}}{\partial x_{2}}=\frac{\partial v_{2}}{\partial x_{1}}
 \end{equation}
 
 \noindent
 we finally obtain
 
 \begin{equation}
     \frac{\partial ^{2}\rho}{\partial t ^{2}}=\left[\frac{\partial ^{2}(\rho v^{2}_{1})}{\partial x_{1}^{2}}+\frac{\rho}{m_{1}}\frac{\partial ^{2}(V+Q)}{\partial x_{1}^{2}} +(1 \leftrightarrow 2)\right]+ 2\frac{\partial ^{2}(\rho v_{1}v_{2}) }{\partial x_{1}\partial x_{2}}
 \end{equation}
 
 \noindent
 or
 
  \begin{equation}
     \frac{\partial ^{2}\rho}{\partial t ^{2}}=-\frac{1}{\hbar ^{2}}\left[\hat{p}^{2}_{1}(\rho v^{2}_{1})+\frac{\rho}{m_{1}}\hat{p}^{2}_{1}(V+Q)+(1\leftrightarrow 2)\right]-\frac{2}{\hbar ^{2}}\hat{p}_{1}\hat{p}_{2}(\rho v_{1}v_{2}).
 \end{equation}
 
 \noindent
 The last terms in (46) or (47) indicate the non-separability of the biparticle compound. They exhibit particular conceptual interest as the operators act at two distant sites and cannot be ascribed to one or the other individual particle. The interparticle expressions for the many-body case follow this same pattern.
 
 \noindent
 Using index notation we write two-particle analogue of (26), setting $\varepsilon=0$, as
 
\begin{equation}
   \frac{\partial ^{2}\rho}{\partial t ^{2}}=\frac{\partial ^{2}T^{(1)}_{ij}}{\partial x^{(1)}_{i}\partial x^{(1)}_{j}}+\frac{\partial ^{2}T^{(2)}_{ij}}{\partial x^{(2)}_{i}\partial x^{(2)}_{j}}+2\frac{\partial ^{2}\left(\rho v^{(1)}_{i}v^{(2)}_{j}\right)}{\partial x^{(1)}_{i}\partial x^{(2)}_{j}}
   \end{equation}

 \begin{equation}
T^{(k)}_{ij}=\rho v^{k}_{i}v^{k}_{j}-\frac{\rho}{m} V\delta _{ij}+\sigma_{ij}, k=1,2
   \end{equation}
   
 \noindent
(upper indices denote number of particle and lower the velocity field components.)

\end{document}